\documentclass[twocolumn,aps,altaffilletter]{revtex4}
\usepackage{graphicx}
\usepackage{color}
\usepackage[colorlinks=false]{hyperref}
\usepackage{bm}
\usepackage{xspace}
\usepackage{times}
\usepackage{graphics}

\newcommand{\units}[1]{\ensuremath{\,{\rm #1}}}

\begin{document}

\title{Quadratic Mixing of Radio Frequency Signals using Superconducting
Quantum Interference Filters}
\author{P.~Caputo}
\author{J.~Tomes}
\author{J.~Oppenl{\"a}nder}
\author{Ch.~H{\"a}ussler}
\author{A.~Friesch}
\author{T.~Tr\"auble}
\author{N.~Schopohl}
\email{nils.schopohl@uni-tuebingen.de}
\affiliation{Lehrstuhl f\"ur Theoretische Festk{\"o}rperphysik, Universit{\"a}t T{\"u}%
bingen\\
Auf der Morgenstelle 14, 72076 T{\"u}bingen (Germany) }
\date{\today }

\begin{abstract}
The authors demonstrate quadratic mixing of weak time harmonic electromagnetic fields
applied to Superconducting Quantum Interference Filters,
manufactured from high-$T_{\mathrm{c}}$ grain boundary Josephson junctions
and operated in active microcooler. The authors use the parabolic shape of the dip in
the dc-voltage output around $B=0$ to mix \emph{quadratically} two external
rf-signals, at frequencies $f_{\mathrm{1}}$ and $f_{\mathrm{2}}$ well below
the Josephson frequency $f_{\mathrm{J}}$, and detect the corresponding
mixing signal at $\left\vert {f_{1}-f_{2}}\right\vert $. Quadratic mixing
takes also place when the SQIF is operated without magnetic shield. The
experimental results are well described by a simple analytical model based
on the adiabatic approximation.
\end{abstract}

\maketitle



\begin{figure}[!tb]
\centering
\includegraphics*{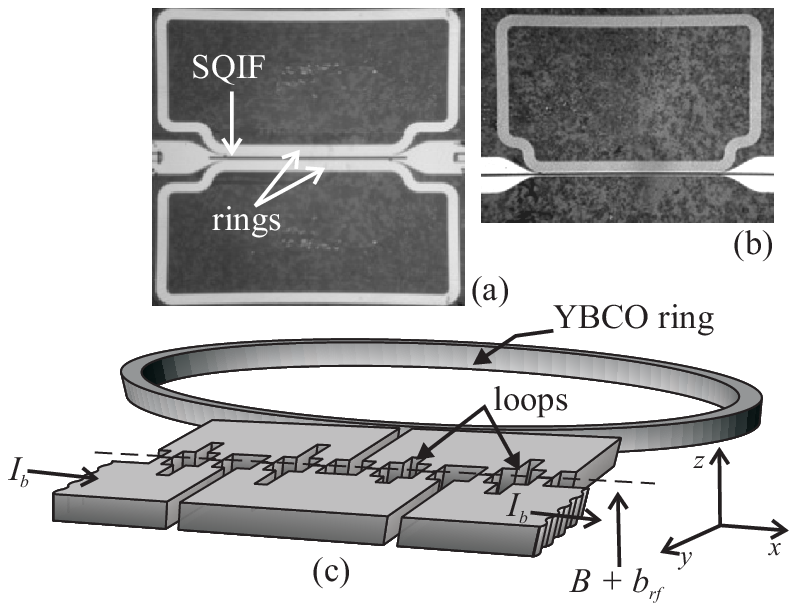}\\
\caption{(a) Optic microscope image of the chip with
the SQIF at the center of the substrate, and the two rings, inductively
coupled to the SQIF; (b) SQIF sensor with one ring. All superconducting
parts, except the regions across the junctions, are covered by gold. (c) Sketch, not in scale, of the SQIF-sensor. The dashed line indicates the grain boundary.}
\label{chip}
\end{figure}

Superconducting Quantum Interference Filters (SQIFs) are Josephson junction (JJ)
interferometers that sense, when operated in the resistive mode, the
presence of a magnetic field $B$ by transforming it to a characteristic $dc$
voltage output $V_{\rm{dc}}(B)$. Upon sweeping a control current through a coil to generate a
suitable compensation field, the voltage output of a SQIF shows a pronounced
unique dip around zero total static field $B=0$. The dip of a SQIF
is characterized by its width $\Delta B$ and its
voltage swing $\Delta V$ \cite{Oppenlander:PRB,Haussler:JAP2001}. The
detailed shape of the voltage output $V_{\rm{dc}}(B)$ can be engineered by choosing suitable areas of the SQIF loops. 
The performance of a SQIF is only weakly sensitive to the spread of the JJ parameters.
These features make SQIFs interesting for applications with ceramic cuprate superconductors.
The sensitivity of a SQIF in response to an applied magnetic field is
determined by the \emph{voltage-to-magnetic field} transfer factor $%
V_{\rm{B}}=\max (\partial V_{\rm{dc}}/\partial B)$. For example, in a serial
SQIF array with $N$ loops the transfer factor $V_{\rm{B}}$ scales with $N$, but
the voltage noise $\sqrt{S_{\rm{V}}}$ derived from the spectral density $S_{\rm{V}}$
(white noise) scales with $\sqrt{N}$, so that the dynamic range $\Delta V/%
\sqrt{S_{\rm{V}}}$ varies proportional to$\sqrt{N}$ \cite%
{Oppenlander:APL_2003,Oppenlander:ASC_2005}. Employing various flux focusing
structures together with a SQIF it is possible to significantly enhance the
transfer factor\cite{Schultze:SST_2003,Schultze:SST_2005}. A flux focusing structure
integrated together with a SQIF (SQIF-sensor) is sketched in
Fig.\thinspace \ref{chip}(c).

In the resistive state, the working point of the device is set by a bias
current $I_{\rm{b}}$ and also by 
a static field $B$.
The dc voltage output $V_\mathrm{dc}(B)$ of the SQIF is the time average of a fast signal  
$V(B,t) \propto \partial \varphi/\partial t$, where $\varphi$ is the
Josephson phase across the JJ. 
The main harmonic of $V(B,t)$ has frequency $f_J = V_\mathrm{dc}(B)/\Phi_0$ (the second Josephson relation)\cite{BP}. In the presence of additional weak radiofrequency
magnetic field $b_{\rm{rf}}\left( t\right)$ which varies slowly in comparison with $f_{J}$, the SQIF voltage can be written as

\begin{eqnarray}
  &&\left\langle V(B + b_\mathrm{rf}(t),t) \right\rangle = \label{Eq:Vdc(t)}
  \\
  &=&
  V_\mathrm{dc}(B) + V_\mathrm{dc}'(B) \cdot b_\mathrm{rf}(t) 
  +   \frac12 V_\mathrm{dc}''(B) \cdot b_\mathrm{rf}^2(t) + \ldots,  \nonumber
\end{eqnarray}
where prime denotes the derivative with respect to $B$, and $\langle \rangle$ the time average over the time scale set by $f_J^{-1}$, while the slow time dependence remains. 
For $b_\mathrm{rf}(t) = b_1\cos(2\pi f_1 t)$, with amplitude $b_1 \ll \Delta B$ and  
frequency $f_1<\frac{f_J}{2}$, Eq.~(\ref{Eq:Vdc(t)}) describes a superposition
of the dc voltage output $V_\mathrm{dc}(B)$ 
with a slowly varying \textit{rf} voltage output at frequencies $f_1$, $2f_1$, etc. The amplitude of the first harmonic of the Fourier spectrum of the output signal contains information about the slope $V_\mathrm{dc}'(B)$, while the amplitude of the second harmonic is proportional to the curvature $V_\mathrm{dc}''(B)$.

The Fourier transform of $V(B+b_\mathrm{rf}(t),t)$ is the \emph{spectral voltage output} $\widehat{V}(B,f)$. The power spectrum $\left\vert \widehat{V}\left( B,f\right) \right\vert ^{2}$
vs. 
$B$ can be detected at frequencies around the center
frequency $f_{1}$ using a spectrum analyzer. If 
$b_{1}$ $\ll $ $\Delta B$, the spectral voltage output will be proportional to $V_\mathrm{dc}'(B)$: it will be
maximum in field regions of maximum slope, 
while in regions of a flat $V_\mathrm{dc}(B)$ curve, say around $%
B=0$, it will tend to zero.

Quadratic mixing is expected for two tone experiments, where the incident 
\textit{rf} signal is a superposition of two time harmonic signals: $%
b_{\rm {rf}}\left( t\right) =b_{1}\cos \left( 2\pi f_{1}\,t\right) +b_{2}\cos
\left( 2\pi f_{2}\,t\right)$ (we assume $f_{1},f_{2}<\frac{f_{J}}{2}$).
Taking into account the parabolic shape of the $V_\mathrm{dc}(B)$-curve around $%
B=0$, one expects in that region a larger amplitude for the quadratic
mixing signals at frequencies $f_{2}-f_{1}$, $2f_{1}$, $f_{2}+f_{1}$ and $%
2f_{2}$, respectively, compared to the field regions where 
$V_\mathrm{dc}'(B)$ is maximal (zero curvature).

Previously we reported on \textit{dc} experiments with SQIF-sensors in
shielded and also in unshielded active micro-coolers \cite{Caputo:APL_2004,Oppenlander:ASC_2005,Caputo:IEEE_2005}. Active
cooling offers numerous advantages: possibility to choose stable temperature down to $50\,{\rm K}$, quick thermal
cycles, and compact and portable setups \cite{AIM}. Results
obtained in unshielded micro-coolers have never revealed a significant degradation of the SQIF-sensor performance with respect to shielded micro-coolers. 
We believe the
principal reason for this is the large dynamic range of the SQIF's.

\begin{figure*}[!t]
\centering
\includegraphics*{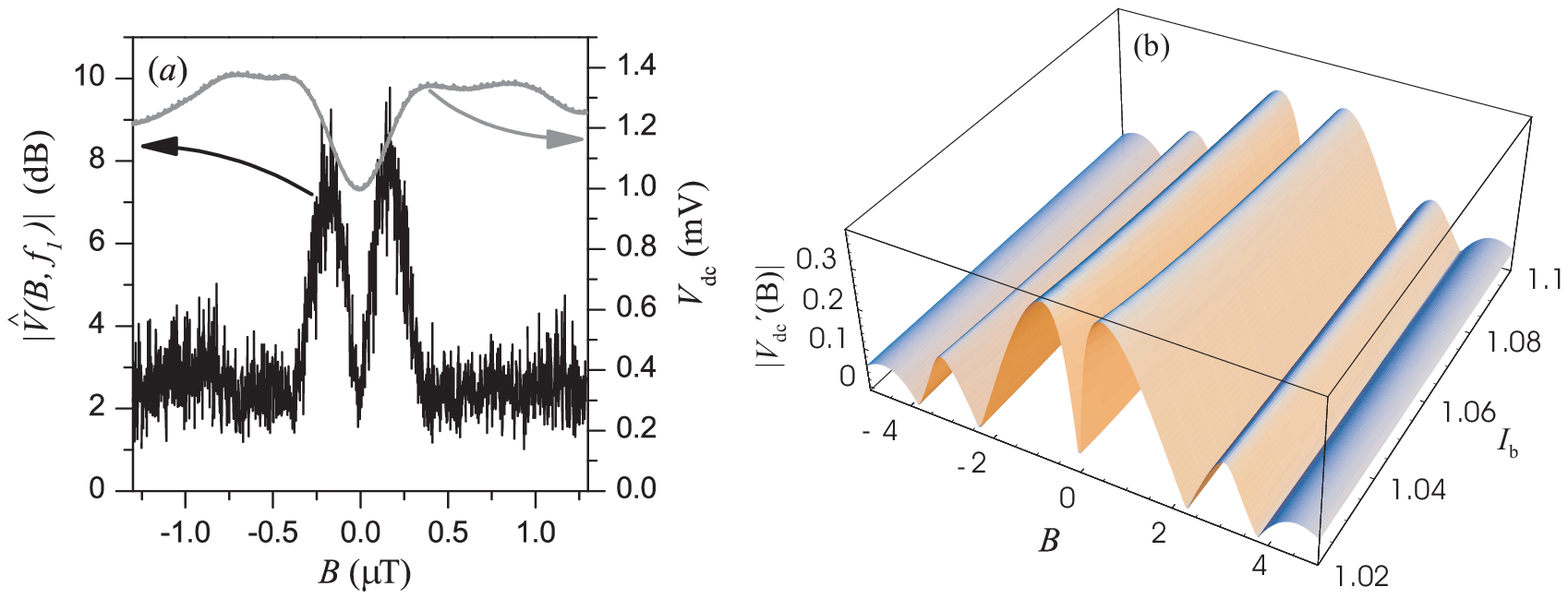}
\caption{(a) SQIF dc voltage $V_{\rm{dc}}$ at $I_{\rm b} = 58\,\mathrm{\protect\mu A}$
and $T= 75\,\mathrm{K}$ vs. static magnetic field $B$ (gray curve, relative to
right axis); simultaneous measurements of the maximum amplitude of the SQIF
\textit{rf} voltage $\left\vert \widehat{V}\left( B, f\right) \right\vert$, induced by the \textit{rf} field with  $f_1=102 \,%
\mathrm{MHz}$ and $b_1=-23 \,\mathrm{dBm}$ (black curve, relative
to left axis). Detection is made by spectrum analyzer in zero span
mode, at $f_1$ and with ResBW= $3 \,\mathrm{kHz}$. The $0\units{dB}$ level is arbitrary. (b) Modulus of the first
derivative of $V_\mathrm{dc}$ vs. static field and bias current.}
\label{plotf1}
\end{figure*}

\begin{figure*}[!t]
\centering
\includegraphics*{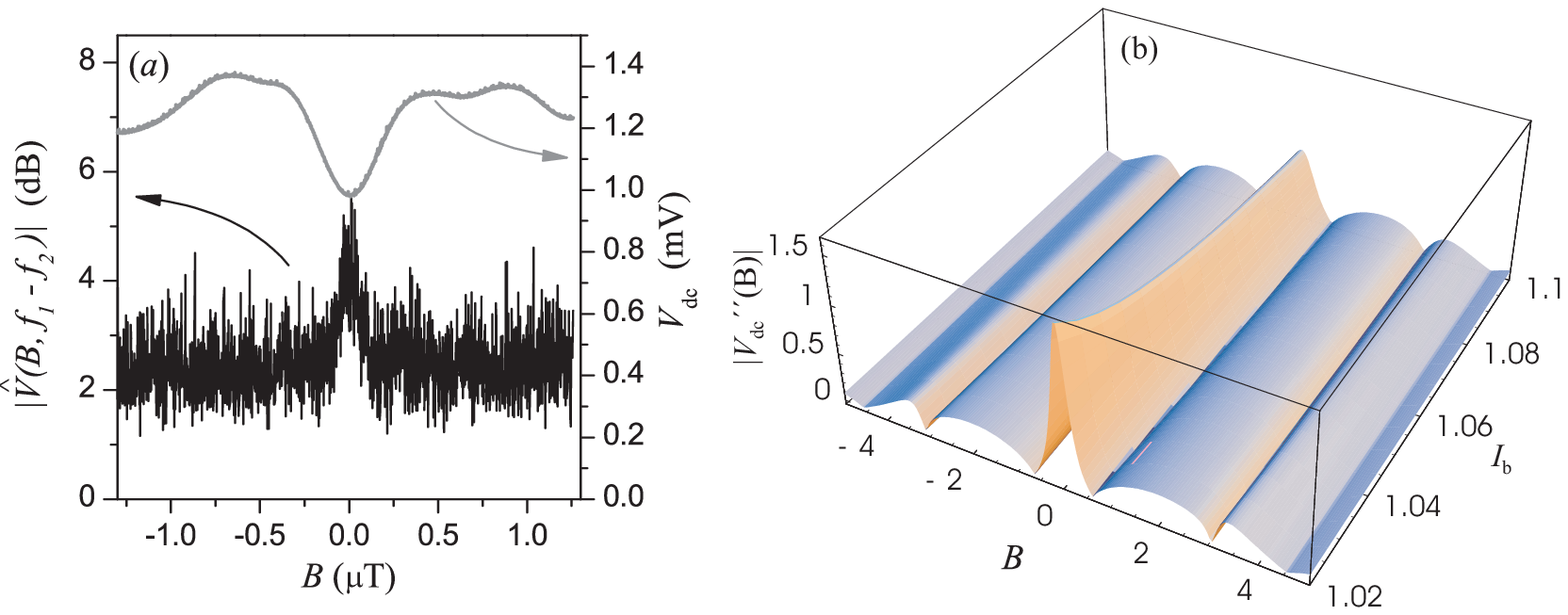}
\caption{(a) Gray curve, relative to right axis: $V_\mathrm{dc}$ vs. $B$; black curve,
rel. to left axis: \textit{rf} voltage $\left\vert \widehat{V}\left( B, f\right) \right\vert$ detected at $\left\vert {f_{1}-f_{2}} \right\vert =120\,\mathrm{MHz}$, with a
ResBW of $3\,\mathrm{kHz}$. The $0\units{dB}$ level is arbitrary. The main signals were at $f_{1}=220\,\mathrm{MHz}$ and $%
f_{2}=100\,\mathrm{MHz}$; amplitudes $b_{1}=b_{2}=-22\,\mathrm{dBm}$. (b) Modulus of the second derivative of $V_{\rm {dc}}$ vs. static field and bias current.}
\label{plotf1f2}
\end{figure*}

In the current experiments we broadcast an \textit{rf}-signal, representing
a superposition of two time harmonic fields, with $f_{1\ }, f_{2}<\frac{f_{J}}{2}$ and small amplitudes $b_{1}$
and $b_{2}$, to a
SQIF-sensor mounted inside the micro-cooler. The measured spectral
voltage output $\left\vert \widehat{V}\left( B,f\right) \right\vert $ of the
SQIF is shown in Figs.\ref{plotf1}--\ref{plotf1f2}. One clearly recognizes besides the
primary signals at frequecies 
$f_{1}$ and
$f_{2}$, a quadratic mixing signal at frequency 
${f_{1}-f_{2}}$, with a maximum amplitude when the SQIF is tuned at the
bottom of the dip at $B=0$. For increasing field $B$, the amplitude of the
quadratic mixing signal symmetrically decreases and apparently vanishes
under the noise floor of the detector as a function of $B$ at approximately
half-way to the position of maximal dip slopes. In sharp contrast to this
behavior, both first order signals 
display 
a maximum amplitude there where the dip slope 
is maximal. The effect is observed for
various SQIF-sensors, with different critical current densities, always with a strenght that increases with the
increase of the transfer factor $V_{B}$. In all
cases, Eq.(1) provides an excellent description of the
detected spectral voltage output vs. static field $B$.


The Josephson junctions in our arrays consist of YBa$_{2}$Cu$_{3}$O$_{7-x}$%
\xspace grain boundary junctions grown on 24$^{\circ }$-oriented bycristal
MgO substrates\cite{IPHT_Jena}. They are designed with a width of $2\,%
\mathrm{\mu m} $, the YBa$_2$Cu$_3 $O$_{7-x}$\xspace layer being $130\,%
\mathrm{nm}$ thick, so that the resulting junction critical current density
is $J_{\rm c} \approx 23 \,\mathrm{kA/cm^2} $, at $T=77\,\mathrm{K}$. The serial
SQIF consists of 211 loops, the distribution of the
loop areas ranging between $38\mathrm{\mu m^{2}}$ and $210\,\mathrm{\mu m^{2}%
}$. In order to vary the coupling of magnetic flux into the SQIF, we
designed two types of focusing structures. Type (a) consists of two
superconducting rings being symmetrically placed on both sides of the grain
boundary [Fig.\thinspace \ref{chip}(a)]. Type (b) is made with only one superconducting ring [Fig.\thinspace \ref{chip}(b)]. Type (b) is effectively a split loop design\cite%
{Ludwig:IEEE2001}, consisting of 10 equidistant parallel thin loops, one
aligned inside the other. The static magnetic field $B$ is applied via a
multi-turn coil placed inside the cooler. The \textit{rf} field $%
b_{\rm {rf}}(t)$ is applied via a $50\,\mathrm{\Omega }$-loop antenna. Experiments
are made either with a mu-metal shield - in this case the \textit{rf}
antenna was placed inside the shield at a distance of about $5\,\mathrm{cm}$
from the chip -, or with an unshielded cryocooler in open space.


We first discuss the experiments with shielded cryo-cooler. We measure the $%
V_{\rm {dc}}(B)$ dependence of the SQIF, and check the dip symmetry with respect
to zero field. There is an optimal temperature at which the voltage swing is
a maximum, at a properly chosen $I_{\rm b}$. At $T\approx 76\,\mathrm{K}$, the
SQIF critical current is $I_{\rm c}\approx 50\mathrm{\mu A}$ and, at $I_{\rm b}=85\,%
\mathrm{\mu A}$, we measure $\Delta V\approx 1160\,\mathrm{\mu V}$ and $%
V_{\rm B}\approx 6500\,\mathrm{V/T}$. The SQIF normal resistance is $%
R=186\,\Omega $. 
Successively, the \textit{rf} is switched on: a time harmonic signal 
is applied to the primary antenna, while the SQIF \textit{rf}
output is detected. 
The value $b_{1}$ is chosen much smaller than
$\Delta B$, to
ensure that the \textit{rf} signal superimposed to the static field does not
modulate the SQIF working point out of the dip and corrupt the effect; but
it has to be high enough so that the spectral voltage output is above the
noise level set by the resolution bandwidth (ResBW) of the spectrum analyzer.

In the single tone \textit{rf }experiment, the incident signal has frequency $f_{1}=102\,\mathrm{MHz}$ and amplitude $%
b_{1}=-23\,\mathrm{dBm}$. At $I_{\rm b}=58\,\mathrm{\mu A}$ and $T=75\,\mathrm{K}
$, the voltage swing is about $400\,\mathrm{\mu V}$. The SQIF rf voltage is amplified by a cold amplifier, designed with high input resistance ($5 \,\mathrm{k\Omega}$).
The spectrum analyzer is operated as a narrowband receiver tuned at
frequency $f\simeq f_{1}$, with a bandwidth set by the ResBW (\emph{zero
span mode}), so that the analog output
corresponds to the maximum signal amplitude. The ResBW is varied from $3\,\mathrm{kHz}$ to $1\,\mathrm{MHz}$. The spectral voltage output of
the SQIF is recorded by computer while slowly sweeping the static field $B$
(sweep frequency in the kHz range). Simultaneously the $V_{\rm {dc}}(B)$ curve is
acquired. Figure \ref{plotf1}(a) displays typical results. Here the $0\units{dB}$ level is arbitrary, being at the present the amplifier gain not exactly known.
It is found that, at frequency $f_{1}$, the spectral voltage
output of the SQIF vs. $B$ is a maximum in field regions
where the dip slopes are maximal. In Fig.\ref{plotf1}(b) we plot the modulus of the first derivative of 
the theoretically calculated voltage output 
vs. magnetic field $B$ and bias current $I_\mathrm{b}$.
%


In the two tone \textit{rf} experiments, the incident field is a linear
combination of two signals with frequency $f_{1}=220\,\mathrm{%
MHz}$ and $f_{2}=100\,\mathrm{MHz}$, and amplitude $b_{1}=b_{2}=-22\,\mathrm{%
dBm}$. In the spectral voltage output of the SQIF, we find then a
quadratic mixing signal at the difference frequency $f=f_{1}-f_{2}=120\,%
\mathrm{MHz}$. The amplitude of this signal is maximal at $B=0$. Figure %
\ref{plotf1f2}(a) displays the spectral voltage output detected at $f=120\,\mathrm{%
MHz}$, with ResBW equal to $3\,\mathrm{kHz}$. Symmetrically, at $f=f_{1}+f_{2}=320\,
\mathrm{MHz}$ a similar spectral voltage response is detected. 
In the current set-up, quadratic mixing has been observed up to few GHz.
In Fig.\ref{plotf1f2}(b) we plot the theoretically calculated $\left\vert V_\mathrm{dc}''(B) \right\vert$ vs. $B$ and vs. $I_{b}$. 
%

In the \textit{rf} experiments without magnetic
shield, the environmental disturbances do not suppress the dip
(although in this case the compensation field is different), and the
quadratic mixing effect as well as the second harmonic generation take
place, similar to the results obtained with the shielded cooler.
In the single tone experiment with a time harmonic \textit{rf} signal at frequency $f_{1}$, we
also detected second harmonic signal generation at $2f_{1}$ in the
spectral voltage output of the SQIF-sensor, with a field dependence around $%
B=0$ as shown in Fig.\ref{plotf1f2}(a).

In conclusion, we have demonstrated that a SQIF-sensor is capable to
transfer the modulations of an incident time harmonic electromagnetic signal
with carrier frequency $f_1<\frac{f_{J}}{2}$ into a corresponding 
\textit{rf }voltage output. 
This suggests potential applications of a SQIF as a
non-linear mixing device, capable to operate at frequencies from dc to few
GHz with a large dynamic range. All experiments have shown that the strength of
the \textit{rf} voltage output at frequency $f_{1}$ depends crucially on
the slope of the $V_{\rm {dc}}(B)$ curve: SQIF-sensors with 
smaller transfer factor $V_{\rm {B}}$ have a reduced maximum amplitude in their 
\textit{rf }voltage output.

The authors are most grateful to R.~IJsselsteijn to provide us with samples
of various critical current densities, and thank V.~Schultze for sharing his insight and 
many useful discussions.

\end{document}